# Alignment of the stellar spin with the orbits of a three-planet system

Roberto Sanchis-Ojeda[1], Daniel C. Fabrycky[2], Joshua N. Winn[1], Thomas Barclay[3,4], Bruce D. Clarke[4,5], Eric B. Ford[6], Jonathan J. Fortney[2], John C. Geary[7], Matthew J. Holman[7], Andrew W. Howard[8], Jon M. Jenkins[4,5], David Koch[4], Jack J. Lissauer[4], Geoffrey W. Marcy[8], Fergal Mullally[4,5], Darin Ragozzine[7], Shawn E. Seader[4,5], Martin Still[3,4] & Susan E. Thompson[4,5]

[1]Department of Physics, Massachusetts Institute of Technology, 77 Massachusetts Ave., Cambridge, MA 02139, USA.

[2]Department of Astronomy and Astrophysics, University of California, Santa Cruz, CA 95064, USA.

[3]Bay Area Environmental Research Institute, 560 Third St. West, Sonoma, CA 95476, USA.

[4]NASA Ames Research Center, Moffett Field, CA 94035, USA.

[5]SETI Institute, 189 Bernardo Ave. #100, Mountain View, USA.

[6]University of Florida, 211 Bryant Space Science Center, Gainesville, Florida 32611-2055, USA.

[7]Harvard-Smithsonian Center for Astrophysics, 60 Garden Street, Cambridge, MA 02138, USA.

[8]Department of Astronomy, University of California, Berkeley, CA 94720, USA.

**The Sun's equator and the planets' orbital planes are nearly aligned, which is presumably a consequence of their formation from a single spinning gaseous disk. For exoplanetary systems this well-aligned configuration is not guaranteed: dynamical interactions may tilt planetary orbits, or stars may be misaligned with the protoplanetary disk through chaotic accretion[1], magnetic interactions[2] or torques from neighbouring stars. Indeed, isolated 'hot Jupiters' are often misaligned and even orbiting retrograde[3,4]. Here we report an analysis of transits of planets over starspots[5–7] on the Sun-like star Kepler-30 (ref. 8), and show that the orbits of its three planets are aligned with the stellar equator. Furthermore, the orbits are aligned with one another to within a few degrees. This configuration is similar to that of our Solar System, and contrasts with the isolated hot Jupiters. The orderly alignment seen in the Kepler-30 system suggests that high obliquities are confined to systems that experienced disruptive dynamical interactions. Should this be corroborated by observations of other coplanar multi-planet systems, then star–disk misalignments would be ruled out as the explanation for the high obliquities of hot Jupiters, and dynamical interactions would be implicated as the origin of hot Jupiters.**

Kepler-30 is a star of nearly solar mass and radius, but it is probably younger than the Sun, judging from its faster rotation and more prominent starspots[8]. The starspots are crucial to measuring the stellar obliquity (the angle between the rotational and orbital angular momentum vectors). Starspots produce two effects: quasi-periodic variation

(QPV) in flux caused by rotation, and shorter-term "anomalies" in flux caused by the transit of a planet in front of a spot. The obliquity can be measured if one observes a sequence of anomalies[5,7], or a few single anomalies and the accompanying QPV[6], as long as the effects of a single spot or compact group of spots can be isolated. This technique has been previously applied to solitary short-period planets, but not longer-period planets or systems of multiple planets. The other widely used technique for measuring stellar obliquities, the Rossiter-McLaughlin effect[9], relies on precise spectroscopy during transits and would be impractical for a star as faint as Kepler-30.

We analyzed 2.5 years of nearly continuous photometric time-series data from the *Kepler* space telescope[10]. The dataset includes 27 transits of Kepler-30b ('planet b'; orbital period, ≈ 29 days, radius, ≈ 4 $R_E$, where $R_E$ is the Earth's radius), 12 transits of Kepler-30c ('planet c'; 60 days; 13 $R_E$), and 5 transits of Kepler-30d ('planet d'; 143 days; 10 $R_E$). After removing instrumental artifacts (see Supplementary Information) we detected QPV with an amplitude (peak-to-peak) of 1.5%. The stellar rotation period is 16.0 ± 0.4 days based on a Lomb-Scargle periodogram[11] (Supplementary Information).

To enable the obliquity analysis we searched for anomalies during transits that are large enough in amplitude and long enough in duration to be caused by the same spots that produce the QPV. Many such anomalies were identified during transits of the largest planet, c. A strong correlation exists between the timing of the anomaly relative to mid-transit, and the phase of the QPV: anomalies observed near mid-transit are found when the QPV is near a local minimum, while anomalies occurring before (or after) mid-transit are found before (or after, respectively) a local minimum. This is the signature of a low-obliquity star[6].

We used both of the above-mentioned methods to establish the quantitative bounds on the obliquity: (1) quantifying the relationship between the anomalies and QPV; and (2) modeling a particular pair of transits for which the anomalies can be attributed to transits over the same spot. To support both of these methods we determined the basic transit parameters -such as the planet-to-star radius ratio ($R_{pl}/R_{star}$) and impact parameter- by fitting the transit data with a standard model for the loss of light during a planetary transit[12]. We excluded the anomalies from the fit, and accounted for transit depth variations due to unoccluded spots (see Supplementary Information, and Supplementary Figure 1). Results are given in Table 1.

The premise of the first method is that any spot that causes an anomaly must also contribute to the QPV. For a given spatial orientation of the star, geometry dictates a specific relationship between the timing of the anomaly and the phase of the QPV. However, all spots contribute to the QPV, not just the occulted spot. Therefore, to measure the obliquity, we must associate each anomaly with a particular component of the QPV. Out of concern that such associations are ambiguous, we exhaustively tried all plausible associations. We rank-ordered the anomalies in order of the loss of light produced by the spot, and focused attention on the five strongest anomalies. We measured the time of each anomaly relative to mid-transit, as well as the time of the transit relative to each local minimum in the QPV within a rotation period (see Supplementary Information). For one of the anomalies there is only one plausible choice for the associated local minimum, while in each of the other 4 cases there are two candidate local minima, giving a set of 16 possible associations. We find that only one of these 16 is compatible with a single orientation of the host star, and in that case the stellar equator is aligned on the sky with the planet's orbit (see Fig. 1, and Supplementary Figure 2). We explored all allowed orientations with a Monte Carlo Markov Chain (MCMC) algorithm[13], finding the sky-projected obliquity to be 4° ± 10°.

For the second method, we searched for pairs of anomalies produced by the same spot. Between successive transits of planet c, a spot will rotate 3.77 times around the star, thereby advancing in longitude by 0.77 of a full circle or 277°, relative to the meridian defined by the sky projection of the stellar rotation axis. An advance by 277° is equivalent to regression by 83°. Therefore, if a spot persists for at least 4 rotations, and if the spot's trajectory is parallel to the planet's trajectory (i.e. if the obliquity is low), then an anomaly observed in the second half of a transit should be followed by an anomaly in the first half of the next transit. The two anomalies should differ by 83° in the suitably defined "anomaly phase" (see Fig. 1).

Two of the five strongest anomalies have this expected phase relationship (see Figure 2), corroborating the finding of a low obliquity. The QPV produced by this spot is coherent over the interval spanned by the two transits, confirming the persistence of the spot (see Supplementary Table 2). Figure 2 shows a spot model fitted to the transit data. For completeness, three spots were included in the model, although only the largest spot (labelled 1) bears information on the stellar obliquity, since it was transited twice by planet c. The model parameters include the spin orientation of the star, the rotation period, and the spot properties (sizes, locations, and intensities). Because the rotation period and spot properties are constrained externally from the QPV, the model could be used to constrain the spin orientation, with results given in Table 1 (see Supplementary Information for details), including a sky-projected obliquity –1 ± 10 degrees. This low sky-projected obliquity is likely to be representative of the true obliquity[14].

Furthermore, all three planetary orbits must be nearly coplanar. The mere existence of multiple transiting planets suggests coplanarity[15], although the possibility remains that the orbits are mutually inclined with nodes (lines of intersection) that happen to lie along the line of sight. However, for Kepler-30, such mutual inclinations would be detectable through variations in transit times and durations caused by nodal precession. To quantify this argument, we performed a 4-body integration of Newton's equations[16-18]. To be compatible with the observed transit times and durations, the mutual inclinations must be smaller than a few degrees. A by-product of our dynamical analysis combined with the transit analysis and the known mass of the star[7] is the determination of the planetary masses and radius (Fig. 3, Table 1).

Such an orderly arrangement might seem to be a natural consequence of the standard model of planet formation, based on core accretion within a flat disk[19]. Recently, though, the host stars of many "hot Jupiter" systems have been found with high obliquities, in some cases even spinning backward relative to the planetary orbit[3,4]. Indeed, it has been argued that stars with hot Jupiters had initially random obliquities, and the only reason low obliquities are sometimes observed is the obliquity-damping effect of planet-star tidal interactions[4].

The observed high obliquities in hot-Jupiter systems have been interpreted as evidence that hot Jupiters attained their close-in orbits through dynamical interactions (which can strongly perturb a planet's orbital orientation) followed by tidal capture. This view is in opposition to the previous paradigm for the origin of hot Jupiters, in which a gradual transfer of energy and angular momentum to the protoplanetary disk causes their orbits to shrink (and maintain a fixed orientation).

One reason why this scenario involving dynamical and tidal interactions has not gained universal acceptance is that obliquity measurements were previously confined to giant planets with small periastron distances. One would like to make sure that the high obliquities are indeed confined to systems that have experienced dynamical interactions. Otherwise it remains possible that stars and their disks are generally misaligned for reasons unrelated to planets, such as chaotic accretion[1], magnetic interactions[2] or differential torques produced by a neighboring star.

Kepler-30 is the type of system that needed to be checked: the coplanarity of the planetary orbits suggests a quiescent history without disruptive dynamical interactions, and the planets are too far from the star for strong tidal interactions. The system was selected by virtue of significant spot-crossing anomalies, and not by any criterion that would have biased the result toward low obliquity. Therefore the observed low obliquity is a clue that the primordial stellar misalignments are not the correct explanation for the high obliquities of hot Jupiter hosts, and that hot Jupiters arise from dynamics and tidal capture. There is only a 6% chance of observing such a low obliquity for Kepler-30 if obliquities were drawn from a random initial distribution. To strengthen our interpretation, additional observations of coplanar multiple-planet system are warranted, and are predicted to yield low obliquities.

1. Bate, M. R., Lodato, G., & Pringle, J. E. Chaotic star formation and the alignment of stellar rotation with disc and planetary orbital axes. *Mon. Not. R. Astron. Soc.* **401**, 1505-1513 (2010).
2. Lai, D., Foucart, F. & Lin, D. N. C. Evolution of spin direction of accreting magnetic protostars and spin-orbit misalignment in exoplanetary systems. *Mon. Not. R. Astron. Soc.* **412**, 2790-2798 (2011).
3. Triaud, A. *et al*. Spin-orbit angle measurements for six southern transiting planets. New insights into the dynamical origins of Hot Jupiters. *Astron. & Astrophys.* **524**, 25-46 (2010).
4. Winn, J. N., Fabrycky, D., Albrecht, S., & Johnson, J. A. Hot Stars with Hot Jupiters Have High Obliquities. *Astrophys. J.* **718**, L145-L149 (2010).
5. Sanchis-Ojeda, R. *et al*. Starspots and Spin-Orbit Alignment in the WASP-4 Exoplanetary System. *Astrophys. J.* **733**, 127-135 (2011).

6. Nutzman, P. A., Fabrycky, D. C., & Fortney, J. J. Using Star Spots to Measure the Spin-orbit Alignment of Transiting Planets. *Astrophys. J.* **740**, L10-L14 (2011).
7. Désert, J. M. *et al*. The Hot-Jupiter Kepler-17b: Discovery, Obliquity from Stroboscopic Starspots, and Atmospheric Characterization. *Astrophys. J. Suppl.* **197**, 14-26 (2011).
8. Fabrycky, D. C. *et al*. Transit Timing Observations from Kepler: IV. Confirmation of 4 Multiple Planet Systems by Simple Physical Models. *Astrophys. J.* **750,** 114-130 (2012).
9. Winn, J. N. *et al*. Measurement of Spin-Orbit Alignment in an Extrasolar Planetary System. *Astrophys. J.* **631**, 1215-1226 (2005).
10. Borucki, W. J. *et al*. Kepler planet-detection mission: introduction and first results. *Science* **327**, 977–980 (2010).
11. Scargle, J. D. Studies in astronomical time series analysis. II - Statistical aspects of spectral analysis of unevenly spaced data. *Astrophys. J.* **263**, 835S-853S (1982).
12. Mandel, K., & Agol, E. Analytic Light Curves for Planetary Transit Searches. *Astrophys. J.* **580**, L171-L175 (2002).
13. Ford, E. Improving the Efficiency of Markov Chain Monte Carlo for Analyzing the Orbits of Extrasolar Planets. *Astrophys. J.* **642**, 505-522 (2006).
14. Fabrycky, D. C. & Winn, J. N. Exoplanetary Spin-Orbit Alignment: Results from the Ensemble of Rossiter-McLaughlin Observations. *Astrophys. J.* **696**, 1230-1240 (2009).
15. Lissauer, J. *et al*. Architecture and Dynamics of Kepler's Candidate Multiple Transiting Planet Systems. *Astrophys. J. Suppl.* **197**, 8-33 (2011).
16. Holman, M. J. & Murray, N. W. The use of transit timing to detect terrestrial-mass extrasolar planets. *Science* **307**, 1288–1291 (2005).
17. Agol, E., Steffen, J., Sari, R. & Clarkson, W. On detecting terrestrial planets with timing of giant planet transits. *Mon. Not. R. Astron. Soc.* **359**, 567–579 (2005).
18. Holman, M. J., et al. Kepler-9: a system of multiple planets transiting a sun-like star, confirmed by timing variations. *Science* **330**, 51-54 (2010).
19. Lissauer, J. J. Planet formation. *Annual Rev. of Astron. and Astrophys.* **31**, 129-174 (1993).
20. Fortney, J. J, Marley, M. S., Barnes, J. W. Planetary Radii across Five Orders of Magnitude in Mass and Stellar Insolation: Application to Transits. *Astrophys. J.* **659**, 1661-1672 (2007).

**Supplementary Information** is linked to the online version of the paper at www.nature.com/nature.

**Acknowledgements** Kepler was competitively selected as the tenth Discovery mission. Funding for this mission was provided by NASA's Science Mission Directorate. The data presented in this Letter were obtained from the Mikulski Archive for Space Telescopes (MAST). STScI is operated by the Association of Universities for Research in Astronomy, Inc., under NASA contract NAS5-26555. Support for MAST for non-HST data is provided by the NASA Office of Space Science via grant NNX09AF08G and by other grants and contracts. S. Albrecht, E. Agol, J. A. Carter, L. Doyle and A. Shporer provided comments on the manuscript. D.C.F. acknowledges NASA support through Hubble Fellowship grant HF-51272.01-A, awarded by STScI. D.R. acknowledges the Harvard Institute for Theory and Computation. E.B.F., M.J.H. and J.N.W. acknowledge NASA support through the Kepler Participating Scientist programme.

**Author Contributions** R.S.-O. performed the spot analysis and wrote the first draft of the paper. D.C.F. performed the dynamical analysis, contributed to the spot modelling, and helped write the paper. J.N.W. contributed to the modelling and helped write the paper. The remaining authors listed below contributed equally. T.B., B.D.C., F.M., S.E.S., M.S. and S.E.T. worked on the data collection, processing and review that yielded the time-series photometry. E.B.F. and M.J.H. provided feedback on the text and interpretation. J.J.F. and J.J.L. worked on elucidating the structure and radii of the planets. J.C.G. worked on the development of Kepler spacecraft photometer electronics, is a builder of Keplercam for the Kepler Input Catalog and follow-up spectral typing. A.W.H. contributed the Keck-HIRES spectra from which the stellar properties were derived. J.M.J. is the Co-Investigator for Data Analysis and designed and built the pipeline that produced the light curves on which this paper is based. D.K. contributed to the concept, design, development, testing and commissioning of the Kepler Mission. G.W.M. contributed the Keck spectroscopy and helped with the imaging, and with some other parts of the original follow-up observations. D.R. studied the possibility of mutual transits and provided feedback on the technical details of the analysis. All authors discussed the results and commented on the manuscript.

**Author Information** Reprints and permissions information is available at www.nature.com/reprints. The authors declare no competing financial interests. Readers are welcome to comment on the online version of this article at www.nature.com/nature. Correspondence and requests for materials should be addressed to R.S.-O. (rsanchis86@gmail.com) or D.C.F. (daniel.fabrycky@gmail.com).

## Host star parameters*

| | |
|---|---|
| KIC/KOI number | 3832472 / 806 |
| Kepler magnitude | 15.4 |
| Mass [solar mass] | 0.99 ± 0.08 |
| Radius [solar radius] | 0.95 ± 0.12 |
| Effective temperature [K] | 5498 ± 54 |
| Gyrochronology age estimate [Gyr] | 2.0 ± 0.8 |
| Quadratic LD coeff. $u_1$ | 0.38 ± 0.09 |
| Quadratic LD coeff. $u_2$ | 0.40 ± 0.19 |
| Linear LD coeff. u | 0.54 ± 0.02 |
| Stellar density [g/cm$^3$] | 2.00 ± 0.10 |

## Planetary parameters†

| Parameter [units] | Planet 30b | Planet 30c | Planet 30d |
|---|---|---|---|
| Orbital period [days] | 29.334 ± 0.008 | 60.3231 ± 0.0002 | 143.343 ± 0.009 |
| Mid-transit time [BJD] | 2455246.65 ± 0.04 | 2455357.8870 ± 0.0005 | 2455273.530 ± 0.010 |
| Eccentricity e | 0.042 ± 0.003 | 0.0111 ± 0.0010 | 0.022 ± 0.005 |
| Periapse angle ω [deg.] | -31 ± 7 | -49 ± 6 | -163 ± 7 |
| Nodal angle Ω [deg.] | 0.03 ± 0.17 | 0 (Relative to 30c) | 1.3 ± 0.5 |
| Planetary mass [M$_⊕$] | 11.3 ± 1.4 | 640 ± 50 | 23.1 ± 2.7 |
| $\vert I - 90° \vert$ [deg] | 0.18 ± 0.16 | 0.32 ± 0.03 | 0.16 ± 0.02 |
| Impact parameter | 0.38 (+0.12, -0.20) | 0.40 (+0.04, -0.06) | 0.38 (+0.08, -0.14) |
| $(R_{pl}/R_{star})^2$ | 0.00165 ± 0.00008 | 0.0162 ± 0.0008 | 0.0083 ± 0.0004 |
| Planet density [g/cm$^3$] | 1.02 ± 0.13 | 1.88 ± 0.17 | 0.19 ± 0.02 |
| Planet radius [R$_E$] | 3.9 ± 0.2 | 12.3 ± 0.4 | 8.8 ± 0.5 |

## Starspot parameters and spin-axis orientation‡

| | |
|---|---|
| Spot rotational period [days] | 16.0 ± 0.4 |
| Spot intensity relative to unspotted photosphere | 0.85 ± 0.03 |
| Inferred spot temperature [Kelvins] | 5298 ± 65 |
| Angular radius of spot [degrees] | 21 (+7, -3) |
| Sky-projected obliquity, recurrence method [degrees] | -1 ± 10 |
| Sky-projected obliquity, 5-anomaly method [degrees] | 4 ± 10 |

*Table 1*. **Parameters of the host star Kepler-30, starspots and planets.**

*Most of the host star parameters are obtained from the literature, and are based on the analysis of high-resolution spectra in conjunction with stellar-evolutionary models[8]. The limb darkening (LD) coefficients are obtained from the light curve analysis (see Supplementary Information). The stellar density is obtained from the dynamical modelling of transit timings and durations.

†Most of the planet parameters are obtained from the four-body dynamical model (see Fig. 3, Supplementary Table 4), with the exceptions of the impact parameters and $(R_{pl}/R_{star})^2$, which are obtained strictly from the light curve analysis. Periods and epochs are best-fits to constant-period models, with error bars reflecting the 1 s.d. spread in the transit timing measurements. $|I - 90°|$ is the deviation of the orbital inclination $I$ from 90° (edge-on). The results for the planetary masses and radii take into account the uncertainty in the assumed stellar mass. The results for $(R_{pl}/R_{star})^2$ are assigned a relative error of 5% to account for possible contamination of the Kepler photometric aperture by background stars. The mass and radius of planet c agree with theoretical models of gas giant planets[20] (see Supplementary Information).

‡The spot parameters are obtained from the spot model (see Fig. 2). In all cases the quoted results and statistical uncertainties are based on the 15.85%, 50% and 84.15% levels of the cumulative a posteriori probability distribution (marginalizing over all other parameters), as determined with the MCMC algorithm.

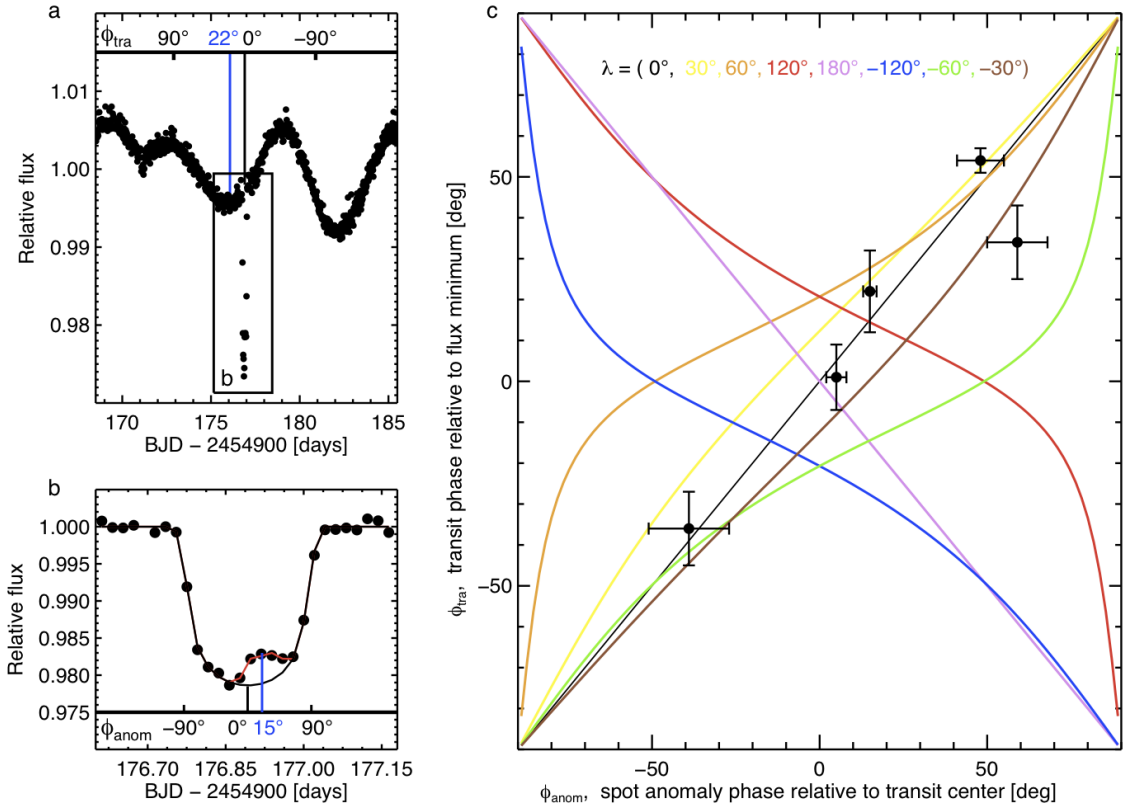

**Figure 1: Evidence for a low obliquity based on transits over several starspots at differing stellar longitudes.**

**a**, A portion of the Kepler light curve, with a box highlighting a transit of Kepler-30c show in panel **b**. The transit occurred just after a local minimum in the QPV. The time of the transit is measured with respect to the selected flux minimum, divided by the rotation period and expressed in degrees, giving the "transit phase" $\phi_{tra} = 22° \pm 10°$. The transit phase is also computed relative to all other local minima within one rotation period. BJD, barycentric Julian day. **b**, A flux anomaly is observed during the transit. The black line is a model without starspots and the red line is a model with one spot. The "anomaly phase", which can be directly compared to the transit phase, is defined by $\sin \phi_{anom} = 2x/L$, where $x$ is the distance from the spot to the center of the transit chord, and $L$ is length of the transit chord. In this case $\phi_{anom} = 15° \pm 2°$, in agreement with $\phi_{tra}$ and consistent with a low obliquity. **c**, Colored lines show the expected relation between $\phi_{anom}$ and $\phi_{tra}$, for different orientations of the star. Since the association between anomalies and minima may be ambiguous, $\phi_{tra}$ was computed for all plausible associations, for the five largest spot anomalies. Only one such set of associations is consistent with a single choice of the stellar orientation. Shown here for that unique choice of associations (see table 3S) is the observed relation between $\phi_{anom}$ and $\phi_{tra}$ implying a projected obliquity $\lambda = 4° \pm 10°$. This error, and the errors on all phases, is $\pm 1$ s.d.

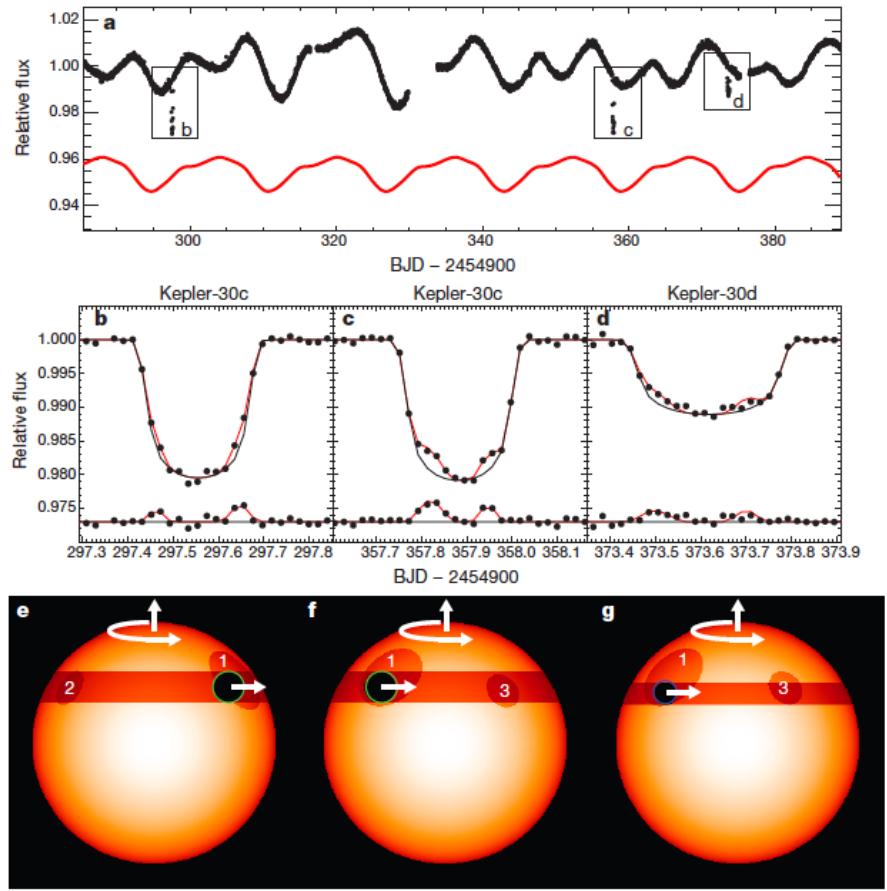

*Figure 2.* **Evidence for a low obliquity, based on a consecutive pair of transits over a single starspot.**

**a,** Data points (black) are a portion of the Kepler light curve, showing the QPV with an approximate 16-day period. The red curve is a model consisting of three spots (shifted vertically for clarity). The model does not take into account spot evolution or differential rotation and is not expected to fit perfectly. Three particular transits are highlighted with boxes and labelled for subsequent discussion. **b,** Light curve of a transit of planet c. The solid dots are data points, the black curve is a transit model with no spots, and the red curve is the best-fitting model with three spots. Residuals from the best-fitting model are displayed near the bottom of the plot. **c,** Same as the previous panel, but for the next transit of planet c. **d,** Same as the previous panel, but for the next transit of planet d. The key parameter of the model, the projected obliquity, was constrained to be smaller than 10°. **e,** Illustration of the stellar disk, dark spots and transit chord for the time range plotted in panel **b**. The white arrows convey the direction of stellar rotation. The black disk represents the transiting planet. **f, g,** Same as panel **e**, but for the time ranges plotted in panels **c, d** respectively. Panels **e** and **f** show that spot 1 was twice eclipsed by planet c, with nearly four stellar rotation periods between the transits. Then, one stellar rotation later, spot 1 was also eclipsed by planet d (panel **g**). (The smaller spot, 3, may also have been eclipsed by both planets during this time interval, though the eclipse by planet d is not securely detected.)

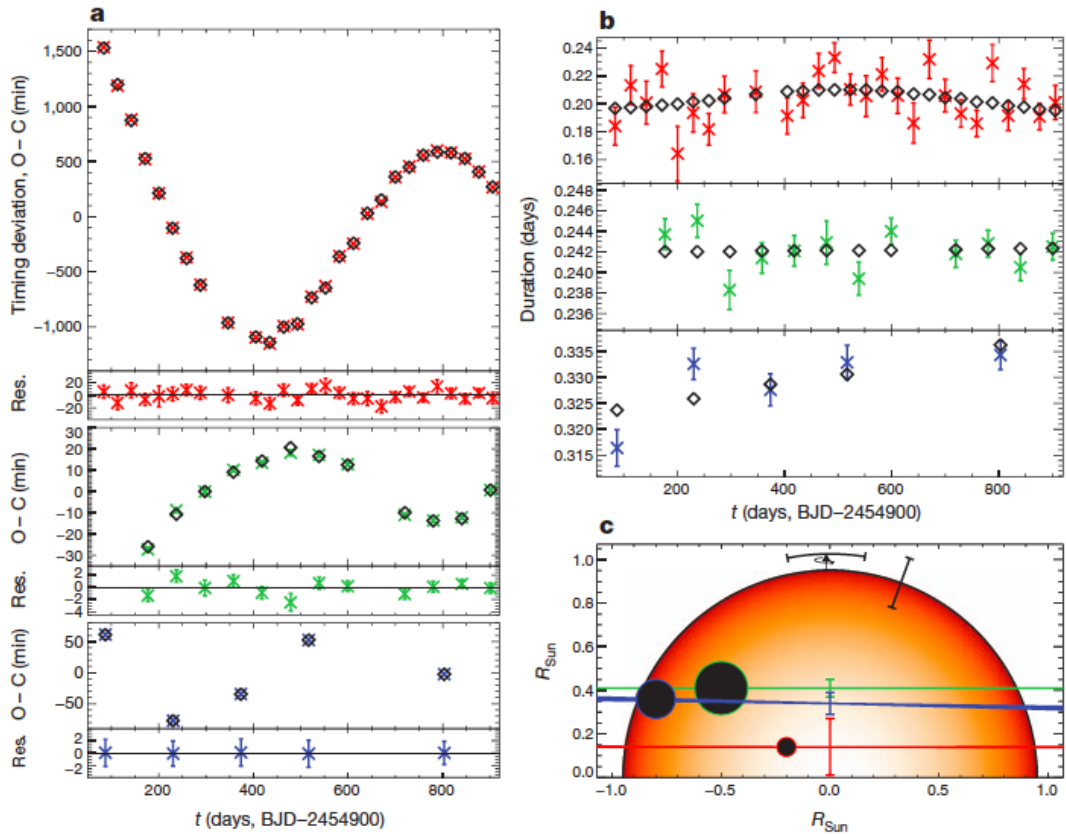

**Figure 3. Evidence for coplanar planetary orbits based on an analysis of transit times and durations.** Throughout, sub-panels and diagram components using colours red, green, and blue refer to planets Kepler-30 b, c, and d, respectively. **a**, The deviation of individual observed (O) transit times (see Supplementary Table 1 with all errors defined as ±1 s.d.) from a constant-period calculation (C) versus time. For planets c and d, suspected starspot-crossing anomalies were masked out before the analysis. Calculated transit times based on a model including planet–planet dynamical interactions[8,18] (Supplementary Information) are shown as open diamonds. Residuals (Res.) between the data and the model are shown below each O _ C plot. **b**, Observed and modelled transit durations. Here the transit duration is defined as the length of time when the centre of the planet is projected in front of the stellar disk. **c**, Diagram of the paths of the planets (black circles with coloured rims) across the face of the star. The error bars show the uncertainty in the impact parameters of the orbits, which are constrained from the timescale of ingress and egress. For planets b and d, three lines are shown, delimiting the 1 s.d. region allowed for the rotations around the line of sight, relative to planet c. The lack of secular changes in the durations (**b**) implies coplanarity to within a few degrees. The error bar on the stellar limb (upper right) is the uncertainty in the stellar radius. The stellar spin axis is denoted (upper middle); its projected orientation is determined from the starspot analysis to be aligned with the planets to within 10° (1 s.d.).

# SUPPLEMENTARY INFORMATION

## 1. The Kepler photometric time series

**Basic characteristics.** This work is based on photometric time-series data from the *Kepler* space telescope[10] obtained between 13 May 2009 and 28 September 2011 (*Kepler* quarters 1 through 10). Until 29 September 2011 the observing mode resulted in one photometric measurement every 29.4 min, whereupon the observing mode was changed to produce a time series with a finer sampling of 58.8 sec.

**Removal of artifacts.** We attempted to remove instrumental artifacts as follows. First we separated the transit segments from the rest of the time series. A transit segment was defined as the data obtained during a given transit along with 3 hours of data before the transit, and 3 hours of data after the transit. For the transit segments, instrumental artifacts were well described by a linear function of time. The parameters of this linear model were determined by fitting a straight line to the out-of-transit data. As for the rest of the data, we subtracted the projections between the data vector and the 4 most significant co-trending basis vectors made available by the *Kepler* project[21]. For some time ranges this correction was not applied, because the data had already been corrected by the *Kepler* project using the PDC-MAP algorithm[22,23].

## 2. Stellar rotation period

**Period determination.** To estimate the stellar rotation period, we divided each quarterly time series by its mean, and then computed a Lomb-Scargle periodogram[11] of the entire time series. A clear peak is observed at 16 days. We interpret this peak as the stellar rotation period. This conclusion was corroborated by a visual inspection of the time series, in which there are at least ten clear cases of flux minima with a consistent amplitude separated by 16 days, for intervals as long as a year. Evidently, there are large and long-lived starspots. Some of these groups of flux minima are studied in more detail in the next section. We adopt an uncertainty of 0.4 days in the rotation period, based on the range of periods giving a periodogram power at least one-third as large as the peak power. Thus the stellar rotation period was estimated to be 16.0 ± 0.4 days.

**Gyrochoronology.** The stellar rotation period can be used to estimate the main-sequence age of the star, because Sun-like stars are observed to slow their rotation according to a simple law in which the rotation period is proportional to the inverse of the square root of the age[24]. We used a polynomial relationship[25] between stellar age, rotation, and mass to estimate the age of Kepler-30. The inputs were the rotation period, taken to be a Gaussian random variable with mean 16.0 days and standard deviation 0.4 days, and the stellar mass, taken to be 0.99 solar masses with a standard deviation of 0.08 solar masses. The resulting distribution of stellar ages has a mean of 2 Gyr and standard deviation of 0.8 Gyr, indicating a star younger than the Sun. The uncertainty of 0.8 Gyr reflects only the uncertainties in the rotation period and stellar mass, and not any systematic errors in the polynomial relationship itself.

## 3. Transit light curve analysis

**Overview.** The analysis of the transit data had several steps, to take advantage of the fact that certain model parameters were assumed to have the same values for all transits, while other parameters were allowed to be specific to each transit. The common parameters were determined by constructing and analyzing a composite transit light curve for each planet, the results of which were then used as constraints in the fit to each individual transit light curve. We performed two iterations of this entire process, the second time enforcing an additional constraint that the orbits are nearly circular, based on the results of the dynamical integration described in Section 6 of this supplement.

**Transit model.** In all cases the transit data were fitted with a standard transit model[12] using a quadratic law to describe the stellar limb darkening, with two free parameters for the limb-darkening coefficients. The planet-to-star radius ratio, scaled stellar radius ($R/a$), and the cosine of the orbital inclination ($\cos I$) were additional free parameters. When data with a cadence of 30 minutes is used, we evaluate the model with a fine time sampling and then time-average the model before comparing it to the data[26].

**Spot corrections.** For planets c and d, the signal-to-noise ratio of the transit data was large enough to justify corrections for spot effects. Spot-crossing flux anomalies were visually identified and excluded from the fit (see Figure 1S). To account for the effect of unocculted starspots we added a new parameter ($L_{spot}$) specific to each transit representing the light lost due to spots, defined as

$$F_{corr} = (F_{trans} - L_{spot}) / (1 - L_{spot})$$

where $F_{trans}$ is the standard transit model with no spots, and $F_{corr}$ is the model that is compared to the data[27]. We allow the $L_{spot}$ parameters to vary freely except for the case of the shallowest transit, for which this parameter was held fixed at zero. Thus we assumed that the effect of unocculted spots was minimal for that transit, and indeed the shallowest transits of both planets c and d occur near a local maximum in the relative flux, as expected if our assumption were correct.

**Parameter estimation.** We determined the best-fitting model parameters by minimizing a standard $\chi^2$ function. The weight of each data point was proportional to the square root of the effective exposure time, and the proportionality constants were determined by the condition $\chi_{min}^2 = N_{dof}$ (number of degrees of freedom) for the best model. Construction of composite light curves allows for a drastic reduction in data volume and consequent speed-up of the MCMC algorithm. We assumed that the limb-darkening parameters, radius ratios, and $R/a$ parameters were constant across all transits of a given planet, but that $\cos I$ (and therefore the transit duration) could vary from one transit to the next. To construct composite light curves, the best-fitting values of the midtransit times were used to calculate the time relative to the nearest mid-transit, and the best-fitting $L_{spot}$ parameters were used to correct the data to zero loss loss-of-light due to unocculted spots. The data were then binned in time with a bin size of 5 minutes. The MCMC algorithm was then used to explore the allowed regions for the global parameters (Table 1). The same MCMC algorithm was also used to obtain the individual transit durations and transit midpoints of each event, using constraints on the other parameters based on the analysis of the composite light curves. The results for the transit midpoints and durations were used as inputs to the dynamical model described in Section 6 (see also Table 1S).

**Iteration with dynamical modeling.** There is a well-known relationship between the orbital parameters, transit parameters, and stellar mean density[28], usually described as a relation between the *R/a* parameter and the stellar mean density for an assumed circular orbit. Therefore, in a system with multiple transiting planets, an additional constraint is available on the orbital and transit parameters by requiring the individual planet models to agree on the stellar mean density. This only useful when the orbital eccentricities of the planets are known or bounded strongly. In the first iteration of our transit analysis, the planets' orbital eccentricities were unknown and were therefore analyzed individually with no common linkage based on the stellar mean density. Subsequently, the dynamical modeling described in Section 6 revealed that the orbital eccentricities must be small. After this finding, we performed a second iteration of the entire process: we repeated our transit analysis with constraints on the orbital eccentricities, thereby gaining additional leverage over the transit parameters, and then refined the dynamical model with the improved parameter set. The output orbital eccentricities were consistent with the results of the first iteration, obviating the need for additional iteration. (We note that iterative procedure could have been avoided by directly coupling the light curve model and dynamical model, a technique that has become known as "photodynamics"[29], at the cost of increased computation time.)

**Limb darkening results.** The fitted limb darkening coefficients $u_1 = 0.38 \pm 0.09$ and $u_2 = 0.40 \pm 0.19$ can be compared with tabulated values based on theoretical models of the atmosphere of the host star[30]. According to those models, a Sun-like star with $\log g = 4.5$, $T_{eff} = 5500$ and $Z = 0.2$ (parameters similar to those of Kepler-30) is expected to have limb-darkening coefficients $u_1 = 0.47$ and $u_2 = 0.22$, in agreement with our results.

## 4. Obliquity determination from transits over starspots at differing longitudes

**Identifying significant anomalies: first method.** When a planet transits a spot, the observed flux is higher than when the planet is transiting the brighter unspotted surface of the star. This is what causes the flux anomalies in the transit light curves. To simplify the analysis we wanted to identify those particular anomalies caused by the largest spots, which are expected to produce the most significant modulation of the out-of-transit flux. One can estimate the total flux deficit caused by the spot—or at least of the portion of the spot transited by the planet—by computing the difference between observed and modeled flux during an anomaly, and then multiplying by an appropriate scale factor[31]. However, this will underestimate the effect of spots that are transited near the limb, due to the effect of geometrical foreshortening. For this reason we employ a modified spot metric,

$$\Delta F = [\Sigma(f_{obs} - f_{theo}) \Delta t / \tau] / (1-r^2)^{1/2}$$

where τ is the ingress time of the transiting planet, r is the projected distance from the center of the spot to the center of the star (in units of the stellar radius), Δt is the time spacing between observations, and $f_{obs}$ and $f_{theo}$ are the observed flux and the (spot-free) modeled flux respectively. The sum is evaluated for all data points during the spot anomaly. We ranked all spot anomalies according to this metric, and identified the six most significant spots, for which Δ F > 0.4%.

**Identifying significant anomalies: second method.** As an alternative means of classifying the spot anomalies, we also fitted a parameterized model to the anomaly data. Our spot model is based on the premise of a limb-darkened star with circular starspots[5]. In addition to the usual transit parameters, which were held fixed in this analysis, there were four parameters for each spot (size, relative intensity, and two-dimensional location in the rotating frame of the star). We specify the spot size by the angular radius, defined as the opening angle of the cone that connects the boundary of the circular spot with the stellar center. Since the rotation period is slow enough that the spot does not move appreciably over the duration of a planetary transit, the model coordinates of the spot are assumed to be constant throughout the transit, coinciding with the projected center of the planet at the midpoint of the anomaly. The size and the relative intensity of the spot are free parameters, as are the transit midpoint and out-of-transit flux level, since those latter two parameters are correlated with the spot parameters. The model flux is calculated as the surface integral of the intensity of the visible hemisphere of the star, excluding the area blocked by the planet. The parameters of the best-fitting model are used to estimate the loss of light due to the entire spot, assuming a circular shape. This is in distinction with the first method, which is less model-dependent but gives only the loss of light due to the portion of the spot that was transited by the planet.

Both methods of ranking the spots give agreement on the top six spots. These spots should produce the largest quasi-periodic flux variations outside of transits. The six largest anomalies should each correspond to a flux variation exceeding 1%, which is readily detectable in the Kepler data.

**Associating flux anomalies with nearby local minima in the out-of-transit flux.** Spots cause a modulation in the disk-integrated flux, as they are carried across the disk by stellar rotation. Due to limb darkening, the loss of light due to a particular spot is largest when that spot is closest to the center of the stellar disk. The quasi-periodic variation thereby encodes some information about the location of the spot, which we use in the obliquity determination. For each of the six transits with the most significant anomalies, we search all of the data within one stellar rotation period to identify local flux minima deeper than 0.4%, i.e., deep enough to be caused by the same spot that is the origin of the transit anomaly. This search becomes more complicated if the transits are located close to a large data gap, like safe mode events, since the shape of the flux minima might be compromised. For this reason we discarded one of the transits with one large anomaly that happened close to the beginning of quarter 10. We checked that dropping this anomaly did not affect the conclusions of this paper.

For one of the remaining 5 transits, only one minimum is identified, and we conclude that the spot that caused the flux anomaly is the same that caused the flux minimum. The alternative interpretations are unlikely. For example there could be bright spots (faculae) situated in such a way as to cancel out the loss of light from the dark spot, but such large faculae have never been observed in active stars[32], and no evidence is found for transits over faculae. Another possibility is that two large spots can combine to cause the same effect as one larger spot. This is possible, but in these cases the two spots would necessarily have a similar rotational phase, and thus the computation of transit phases described below would be largely unaffected. In the other 4 large flux anomalies, there were two local minima in the vicinity of the transit. For these we tried all possible associations between flux anomalies and local minima, as described below.

**Computing $\phi_{tra}$, the phase of each transit within a stellar rotation cycle.** For each transit we computed the phase of the transit ($\phi_{tra}$) relative to each of the candidate minima. The phase is defined as the time of the transit, relative to the time of the flux minimum, divided by the rotation period and expressed in degrees. To measure this transit phase we first needed to measure the times of minimum light. This was done by fitting a parabolic function to the data near the minimum. These timings, along with formal statistical uncertainties, can be found in Table 2S.

In the cases where PDC-MAP data were available, we repeated this procedure with both the flux series obtained with our detrending algorithm (fitting the co-trending vectors) and the PDC-MAP flux series. We found differences up to 0.1 days, several times larger than the formal statistical uncertainties. This demonstrates that the times of minimum light are dependent on the details of the detrending algorithm. Therefore, to obtain more robust results, we analyzed not only the local minimum closest in time to the transit, but the entire periodic sequence of local minima that occur within 4 stellar rotation periods of the transit in question. The large spots evidently lasted for several rotation periods, enabling this analysis. The timings of all those minima are also given in Table 2S. We then fitted the times of minimum light for each spot with a linear function of cycle number. The standard deviation of the residuals—which was up to 20 times larger than the formal statistical uncertainty in each time of minimum light—was adopted as a more realistic estimate of the uncertainty of each of the timings. The slope of the line is interpreted as the period of rotation of the given spot, in all cases close to the value $16.0 \pm 0.4$ days established in Section 1.

The transit phase is then defined

$$\phi_{tra}^j = 360° * (t_0 - t^j) / P$$

where P is the rotation period of the spot, $t^j$ represents the time of minimum light, and $t_0$ is the mid-transit time. The uncertainty in this phase ($\delta\phi_{tra}^j$) is obtained by propagating the uncertainties of all the input parameters.

**Computing $\phi_{anom}$, the phase of each anomaly within the transit.** The timing of the spot-crossing anomaly relative to the mid-transit time also bears information about the location of the spot, in this case with respect to the transit chord. Each spot-crossing anomaly was assigned an anomaly phase ($\phi_{anom}$), defined as

$$\phi_{anom} = \sin^{-1}(x/(1-b^2)^{0.5})$$

where x is the location of the spot measured along the transit chord, in units of the stellar radius, and b is the impact parameter of the transit. To determine this phase and its uncertainty ($\delta\phi_{anom}$), we use the spot transit model previously mentioned, in which x is a free parameter. We used an MCMC algorithm to determine the allowed range of this parameter, and then propagate the uncertainty appropriately to obtain $\delta\phi_{anom}$ (see Table 3S).

**Using the relation between $\phi_{tra}$ and $\phi_{anom}$ to determine the obliquity.** Given a certain spin-orbit orientation and a particular impact parameter, there is a one-to-one geometrical relationship between these two phases. Symbolically we write this relationship as

$$\phi_{tra,theo} = f(\lambda, i_s, \phi_{anom}, b)$$

and, for each of the 16 possible associations between flux anomalies and local minima, we define the goodness-of-fit as

$$\chi^2(\lambda, i_s, b, \phi_{anom}, j) = \Sigma[(\phi_{tra,theo} - \phi_{tra}^j)/\delta\phi_{tra}^j]^2 + \Sigma[(\phi_{anom,param} - \phi_{anom})/\delta\phi_{anom}]^2 + [(b - b_c)/\delta b_c]^2$$

where $\lambda$ is the sky-projected stellar obliquity, $i_s$ is the inclination of the stellar rotation axis with respect to the line of sight, the index j ranges over the 16 possible associations, and $b_c$ and $\delta b_c$ are the measured impact parameter of planet c and its associated uncertainty (Table 1). For each of the 16 possible associations, we evaluate the minimum of the $\chi^2$ function in a 2D uniform grid in $\lambda$ and $i_s$, with $\lambda$ ranging from $-180°$ to $+180°$ and $i_s$ ranging from $0°$ to $180°$, with a spacing of less than half degree. With eight parameters and eleven measurements, we have three degrees of freedom. We only find one association that gives an acceptable fit, with a minimum $\chi^2 \approx 5.2$ and a p-value of 0.16. The next best association gives a minimum $\chi^2 \approx 26.5$, with a p-value of 0.000008. This test thereby uniquely determines the associations between flux anomalies within a transit, and nearby minima in the out-of-transit flux (see Table 3S for final value of the phases). Once this is decided, we used an MCMC algorithm to obtain the final value of $\lambda$ and its uncertainty, using the correct association. (As expected $i_s$ is unconstrained by this analysis.)

## 5. Obliquity determination from two transits over a single starspot

A second, independent determination of the obliquity was undertaken, based on the observed recurrence of flux anomalies by the same spot in two different transits. For this task the spot model was changed appropriately. To give an acceptable fit to the light curves it was necessary to include three spots in the model, even though only one of those spots (the one that was transited twice) is of interest. The largest spot, labeled 1 in Figure

2, is the crucial spot that was transited twice by planet b. The smaller spots 2 and 3 were included for completeness but do not have any bearing on the stellar obliquity. These two spots are fixed to the transit chord as previously explained. For simplicity, all the spots were assigned the same intensity, since for spots 2 and 3 this parameter is degenerate with the spot angular radius. More information is available for spot 1 because Kepler-30c transited this spot twice. The model is also modified (relative to the model described in Section 2) to account for the changing position of the spot on the disk of the star. We model the trajectory of the spot with the two angles specifying the stellar orientation, the rotation period of the star, and a particular time when the spot is closest to the center of the star.

The transit data alone would not allow the spot parameters to be determined uniquely, especially because the transits are well separated in time and the spots are large. However, we can apply some crucial constraints on the model based on the analysis of the out-of-transit quasiperiodic flux variations. Specifically, Gaussian priors were imposed on the stellar rotation period, and on the amplitudes and phases of the out-of-transit flux variations implied by the spot locations (Table 2S). To compute the amplitude of the quasi-periodic flux variations for a given set of spot parameters, we used the Dorren model[33], an analytic expression that gives the loss of light from a circular spot of a certain size, brightness contrast and location. This model uses a linear law for the limb darkening profile. We assumed that the limb-darkening law was the same for spots as for the surrounding photosphere. The spots were required to have a lower intensity than the surrounding photosphere, and a maximum angular size of 60° to protect against outlandish solutions. The individual transit times and out-of-transit flux levels were allowed to vary freely. The allowed regions for the parameters were determined with an MCMC algorithm[13], and are given in Table 1. We used the best-fitting (zero obliquity) solution to plot the quasi-periodic flux variations using the same Dorren model, and in Figure 2a the result is plotted in red. The spot model captures the general amplitude of the modulations and the phase of the largest spot, but does not fit perfectly. This was expected, since we are not modeling all the smaller spots that may exist on the surface or trying to fit the quasi periodic flux variations point by point, nor are we taking into account spot evolution or differential rotation.

## 6. Dynamical modeling

**Overview.** A dynamical model was fitted to the observed transit times and durations, in order to determine the planet masses and especially the mutual inclinations between the planetary orbital planes. The model consisted of four spherical bodies (the star and three planets) dynamically interacting according to Newton's equation of motion. This model was advanced, using a root-finding technique[34], to each moment of closest sky-projected separation between each of the planets and the star. This moment is the model mid-transit time. This distance of closest sky-projected separation, in units of stellar radii, is the model impact parameter $b$ (averaged over the transits which are observed). The model transit duration is the width of the star along that transit chord, $2 R_* \sqrt{1-b^2}$, divided by the sky-projected relative velocity of the planet and the star ($v$). These three types of quantities are compared to the measurements (Table 1S), and the $\chi^2$ function (the sum of the squares of the differences between model and data, normalized by the observational errors) is minimized using the Levenberg-Marquardt algorithm[35].

**Model parameters**. The parameter set used in the model are osculating orbital elements in Jacobian coordinates: each planet's orbit is referenced to the center of mass of all bodies on interior orbits, with instantaneous Keplerian orbits defined using the total mass of all interior bodies and that planet. The numerical integrations use Cartesian, astrocentric coordinates (at a common dynamical epoch BJD 2455550), coordinates into which the parameter set is converted prior to the integration. The parameters are orbital period, P; mid-time of a transit near the dynamical epoch, $T_0$; the parameters (e sin ω) and (e cos ω), where e is the eccentricity and ω is the angle between the periastron and the node, the latter being the location the planet passes through the sky plane moving towards the observer; the inclination of the orbital plane with respect to the plane of the sky, $i$; the rotation angle of the node about the line of sight, Ω. Finally, we fit the mass of each planet with respect to the star, $M_p/M_*$. We have used this method previously to fit transit midtimes[8,18,36], and in Table S4, we give the resulting orbital parameters.

**Obtaining the density of the host star**. An additional step of this analysis was to find the density of the star, $\rho_*$. In practice, we fix the stellar mass at 0.99 $M_{Sol}$ and use stellar radius $R_*$ as an additional fit parameter, which we convert to $\rho_*$ using the adopted stellar mass. The rationale of this approach is that under the transformation of masses $M_* \rightarrow \alpha M_*$, Newton's equations have the scaling property of time $t \rightarrow \alpha^{1/2} t$ and of distances/radii $R_* \rightarrow \alpha^{1/3} R_*$ and thus $M_*/R_*^3 \rightarrow M_*/R_*^3$, meaning that photometric data uniquely constrain only densities. While fitting a certain timing dataset, the fit can still be rescaled to various masses and radii. Another way to demonstrate this is to note the dependencies of parameters which together determine the stellar radius: $R_*=D/(2v\sqrt{(1-b^2)})$. The shape of transits determines the parameter $b$ and duration D; they are independent of $M_*$. The sky-projected orbital velocity v comes from the numerical integration. The orbital period is fixed by the observations, so v scales the same way as semi-major axis with stellar mass, i.e. $v \sim M_*^{1/3}$. Thus the inferred $R_*$ scales as $M_*^{1/3}$, so with the integrations assuming a certain $M_*$, what is really being constrained is the stellar density.

**The best fitting model**. For this analysis, the average impact parameters we used were given in Table 1. The resulting goodness-of-fit statistic and number of data points for were $\chi^2/\#$:

Times of planet b :           18.3/ 27
Times of planet c :           12.9 / 12
Times of planet d :           0.02 / 5
Durations of planet b:        39.9 / 27
Durations of planet c:        16.1 / 12
Durations of planet d:        10.5 / 5
Impact parameter of planet b: 1.4 / 1
Impact parameter of planet c: 0.1 / 1

Impact parameter of planet d: 0.03 / 1

The total $\chi^2$ of 99.4 for 70 degrees of freedom is marginally acceptable: according to the chi-squared test, it has a p-value of 0.012. The durations and impact parameter of planet b have high deviations from their measured errors (Table 1, 1S). Kepler 30 b is a special case because its ingress and egress have very low signal-to-noise per transit, so the determination of errors of durations and impact parameter is especially difficult.

**Mutual events**. Note that planets c and d have nearly the same impact parameter, and there is evidence that they cross the same spot. This suggests that if they transited the star at the same time, their disks might intersect, in projection. Such a geometry would lead to a momentary brightening, relative to the two-planet eclipse model, called a mutual event[37]. In the current dataset, no such anomalies exist, and the best-fitting model has no such events spanning ~8 years of data possible from *Kepler*. However, ground-based telescopes may survey this system thereafter[38], presuming the planets have not nodally precessed onto differing transit chords by then.

**Planet parameters**. Although the main motivation for our dynamical analysis was the determination of mutual inclinations, a by-product is the determination of the planetary masses and densities, which were heretofore poorly known. From table 4S, we obtain the planet to star mass ratio that combined with the stellar mass obtained from the spectra (Table 1) gives us the mass of the planets. This same mass ratio, together with the new precise density of the star, and the planet to star radius ratio, allows us to get the densities. Then it is straightforward to obtain the planetary radius from these. We confirm that b is akin to Neptune, and c is a gas giant similar to Jupiter. Planet d has the lowest mean density of any exoplanet smaller than Jupiter[39], although we caution that the mass of planet d is less robustly constrained than the other two planet masses. The constraint on d's mass relies on the analysis of its gravitational pull on c, which is itself engaged in a resonance with b, making the effects difficult to isolate.

To test the robustness of these measurements, we adopt a theoretical stance and assume that the mass and radius of Kepler 30c should conform to theoretical models of giant planets, which are thought to be reliable for cool (not strongly irradiated) giant planets[20]. Thus, the massive giant planet can be used as a reference object, instead of the usual practice of using the star as the only reference object. With an orbital period of 60 days around a Sun-like star, and being so massive, in theory the size of this planet depends chiefly on its age and the composition of the solid core at its center. With the estimate of the age from the rotational period and the mass fixed to 2 Jupiter masses, we estimate the largest size possible as the cool Jupiter with no core and age of 1 Gyr, which is 1.14 times the radius of Jupiter. On the other end, to provide a lower bound on the planet radius, we choose a cool Jupiter with a very large core, 100 times the mass of Earth, and as old as 4.5 Gyrs, giving a size of 0.97 Jupiter radii. Putting these results together, we set a value for the radius of 1.05 ± 0.09 Jupiter radii for Kepler 30c, or what is the same, 11.8 ± 1.0 Earth radii. With this estimate, and the knowledge of the relative sizes of the planets, one can determine the sizes of the smaller planets, whose radii depend strongly on composition and thus are not well constrained by theory. For Kepler 30b we obtain a radius of 3.8 ± 0.3 Earth radii, and for Kepler 30d we obtain a radius of 8.4 ± 0.8 Earth radii. All these values agree with the observed values, showing the robustness of our analysis. Even using this slightly smaller radius for the Kepler 30d, we obtain a density of 0.21 ± 0.07 g/cm$^3$ that is still the lowest among all exoplanets smaller than Jupiter. We emphasize that in this analysis, theoretical models for giant planets influence the planet properties, whereas the original values reported in Table 1, which have smaller uncertainties, are also independent of such models.

# ADDITIONAL REFERENCES

21. http://archive.stsci.edu/kepler/cbv.html
22. Stumpe, M. C. *et al*. Kepler Presearch Data Conditioning I - Architecture and Algorithms for Error Correction in Kepler Light Curves. PASP, submitted [arXiv:1203.1382] (2012).
23. Smith, J. C. *et al*. Kepler Presearch Data Conditioning II - A Bayesian Approach to Systematic Error Correction. PASP, submitted [arXiv:1203.1383] (2012).
24. Skumanich, A. Time Scales for CA II Emission Decay, Rotational Braking, and Lithium Depletion. *Astrophys. J.* **171**, 565S (1972).
25. Schlaufman, K. C. Evidence of Possible Spin-orbit Misalignment Along the Line of Sight in Transiting Exoplanet Systems. *Astrophys. J.* **719**, 602S (2010).
26. Kipping, D. M., Binning is sinning: morphological light-curve distortions due to finite integration time. *Mon. Not. R. Astron. Soc.* **408**, 1758 (2010).
27. Czesla, S., Huber, K.F., Wolter, U., Schröter, S., Schmitt, J.H.M.M. How stellar activity affects the size estimates of extrasolar planets. *Astron. & Astroph.* **505**, 1277 (2009).
28. Winn, J. Exoplanet transits and occultations, in *Exoplanets*, ed. S. Seager, University of Arizona Press, Tucson, Arizona, p. 55-77 (2010).
29. Carter, J. *et al*. KOI-126: A triply eclipsing hierarchical triple with two low-mass stars. *Science* **331**, 562 (2011).
30. Claret, A. Gravity and limb-darkening coefficients for the Kepler, CoRoT, Spitzer, ubvy, UBVRIJHK, and Sloan photometric systems. *Astron. & Astroph.* **529**, 75 (2011).
31. Deming, D. *et al*. Kepler and Ground-based Transits of the Exo-Neptune HAT-P-11b. *Astrophys. J.* **740**, 33 (2011).
32. Foukal, P. What Determines the Relative Areas of Spots and Faculae on Sun-like Stars? *Astrophys. J.* **500**, 958F (1998).
33. Dorren, J. D. A new formulation of the starspot model, and the consequences of starspot structure. *Astrophys. J.,* **320**, 756D (1987).
34. Fabrycky, D. in Exoplanets (ed. Seager, S.) 217-238 (University of Arizona Press, 2010).
35. Markwardt, C. B. Non-Linear Least Squares Fitting in IDL with MPFIT (proc. *Astronomical Data Analysis Software and Systems XVIII*, Quebec, Canada). ASP Conference Series, Vol. **411**, 251-254, eds. D. Bohlender, P. Dowler & D. Durand (Astronomical Society of the Pacific: San Francisco) 2008.
36. Cochran, W. D. *et al*. Kepler-18b, c, and d: A System of Three Planets Confirmed by Transit Timing Variations, Light Curve Validation, Warm-Spitzer Photometry, and Radial Velocity Measurements. *Astrophys. J.* **197**, 7S (2011).
37. Ragozzine, D., Holman, M. J. The Value of Systems with Multiple Tranisting Planets, e-print, arxiv:1006.3727 (2010).
38. Tingley, B., Palle, E., Parviainen, H. et al. Detection of transit timing variations in excess of one hour in the Kepler multi-planet candidate system KOI 806 with the GTC, Astron. & Astrophys., **536**, 9 (2011).

39. Wright, J. T., Fakhouri, O., Marcy, G. W. The Exoplanet Orbit Database. *The Astronomical Soc. Of the Pacific*, **123**, 412 (2011).

**Table 1S. Transit Durations and midpoint times obtained from the transit model.**

The errors are estimated using an MCMC algorithm. The transit durations of each planet are constant within the errors, which is used to constrain the mutual inclinations. The transits are not equally spaced, due to gravitational interactions between the planets. We used this information to constrain the masses and orbits of the planets (see Figure 3).

| Planet | Transit # | Time [BJD-2454900] | Error | Transit Duration [days] | Error |
|---|---|---|---|---|---|
| b | 0 | 83.719 | 0.007 | 0.184 | 0.013 |
|   | 1 | 112.858 | 0.007 | 0.213 | 0.014 |
|   | 2 | 142.027 | 0.008 | 0.201 | 0.015 |
|   | 3 | 171.159 | 0.007 | 0.225 | 0.013 |
|   | 4 | 200.326 | 0.012 | 0.164 | 0.020 |
|   | 5 | 229.490 | 0.008 | 0.193 | 0.014 |
|   | 6 | 258.684 | 0.006 | 0.182 | 0.011 |
|   | 7 | 287.895 | 0.007 | 0.207 | 0.013 |
|   | 9 | 346.419 | 0.008 | 0.209 | 0.015 |
|   | 11 | 405.094 | 0.007 | 0.191 | 0.013 |
|   | 12 | 434.432 | 0.007 | 0.202 | 0.012 |
|   | 13 | 463.924 | 0.007 | 0.224 | 0.013 |
|   | 14 | 493.316 | 0.006 | 0.233 | 0.011 |
|   | 15 | 522.874 | 0.006 | 0.210 | 0.011 |
|   | 16 | 552.316 | 0.008 | 0.205 | 0.015 |
|   | 17 | 581.892 | 0.006 | 0.221 | 0.012 |
|   | 18 | 611.352 | 0.006 | 0.206 | 0.013 |
|   | 19 | 640.923 | 0.008 | 0.186 | 0.014 |
|   | 20 | 670.380 | 0.007 | 0.232 | 0.014 |
|   | 21 | 699.923 | 0.006 | 0.206 | 0.012 |
|   | 22 | 729.366 | 0.005 | 0.193 | 0.010 |
|   | 23 | 758.817 | 0.005 | 0.186 | 0.009 |
|   | 24 | 788.230 | 0.007 | 0.229 | 0.013 |
|   | 25 | 817.599 | 0.006 | 0.191 | 0.010 |
|   | 26 | 846.940 | 0.006 | 0.214 | 0.011 |
|   | 27 | 876.243 | 0.005 | 0.191 | 0.009 |
|   | 28 | 905.525 | 0.006 | 0.201 | 0.012 |
| c | 0 | 176.8927 | 0.0007 | 0.2437 | 0.0015 |
|   | 1 | 237.2268 | 0.0007 | 0.2450 | 0.0016 |
|   | 2 | 297.5542 | 0.0009 | 0.2383 | 0.0019 |
|   | 3 | 357.8826 | 0.0007 | 0.2414 | 0.0015 |
|   | 4 | 418.2062 | 0.0007 | 0.2421 | 0.0015 |
|   | 5 | 478.5308 | 0.0010 | 0.2429 | 0.0021 |
|   | 6 | 538.8514 | 0.0007 | 0.2394 | 0.0016 |
|   | 7 | 599.1696 | 0.0006 | 0.2440 | 0.0013 |
|   | 9 | 719.7957 | 0.0006 | 0.2418 | 0.0013 |
|   | 10 | 780.1152 | 0.0006 | 0.2428 | 0.0013 |
|   | 11 | 840.4375 | 0.0005 | 0.2405 | 0.0013 |
|   | 12 | 900.7677 | 0.0006 | 0.2425 | 0.0013 |
| d | 0  | 87.2631  | 0.0015 | 0.316  | 0.003  |
|   | 1  | 230.3777 | 0.0014 | 0.333  | 0.003  |
|   | 2  | 373.6182 | 0.0015 | 0.328  | 0.003  |
|   | 3  | 516.8893 | 0.0015 | 0.333  | 0.003  |
|   | 5  | 803.2728 | 0.0013 | 0.334  | 0.003  |

**Table 2S. Measured timings for relevant flux minima used to estimate the rotational phases of the spots occulted during transit.**

The flux minima are grouped according to periodicity, and each group represents one large active region or spot. MCMC errors are based in a parabola fit to each flux minima, whereas the final errors used are based on the standard deviation of the residuals of the linear fit to all the timings of a given group. The rotation period and its error are based on that same linear fit.

The nine timings that occur close to one of the five transits that show large spot-crossing events are underlined. Written in bold and enclosed in boxes are the five flux minima uniquely determined (SI).

| Spot group | Epoch | Timing | MCMC error | Final error | Period | Period error |
|---|---|---|---|---|---|---|
| I | 0 | 150.242 | 0.007 | 0.40 | 16.11 | 0.08 |
| I | 2 | <u>182.153</u> | <u>0.010</u> | 0.40 | 16.11 | 0.08 |
| | 3 | 198.270 | 0.013 | | | |
| | 4 | 213.745 | 0.016 | | | |
| | 5 | 230.999 | 0.046 | | | |
| | 6 | 246.851 | 0.034 | | | |
| II | 0 | 144.264 | 0.016 | 0.44 | 16.01 | 0.08 |
| | 1 | 160.129 | 0.011 | | | |
| | **2** | **175.927** | **0.014** | | | |
| | 4 | 209.054 | 0.063 | | | |
| | 5 | 224.423 | 0.016 | | | |
| | 6 | 239.824 | 0.025 | | | |
| III | 0 | 264.863 | 0.021 | 0.40 | 15.94 | 0.06 |
| Spot 1 | 1 | 280.107 | 0.010 | | | |
| See figure 2 | **2** | **296.037** | **0.012** | | | |
| | 3 | 312.369 | 0.008 | | | |
| | 4 | 328.385 | 0.010 | | | |
| | 5 | 344.021 | 0.012 | | | |
| | **6** | **359.480** | **0.015** | | | |
| | 7 | 376.611 | 0.037 | | | |
| IV | 0 | 259.306 | 0.044 | 0.42 | 14.78 | 0.18 |
| | 1 | 273.199 | 0.012 | | | |
| | 2 | 288.296 | 0.020 | | | |
| | <u>3</u> | <u>303.549</u> | <u>0.027</u> | | | |
| V | <u>0</u> | <u>350.727</u> | <u>0.010</u> | 0.13 | 15.67 | 0.02 |
| | 1 | 366.291 | 0.011 | | | |
| | 2 | 382.016 | 0.011 | | | |
| | 3 | 397.571 | 0.012 | | | |
| | 4 | 413.266 | 0.010 | | | |
|     | 5  | 428.724 | 0.009 |      |       |      |
|     | 6  | 444.762 | 0.008 |      |       |      |
|     | 7  | 460.272 | 0.006 |      |       |      |
|     | **8**  | **476.165** | **0.008** |      |       |      |
| VI  | 0  | 681.771 | 0.014 | 0.57 | 15.16 | 0.12 |
|     | 2  | 712.747 | 0.014 |      |       |      |
|     | 3  | 726.574 | 0.069 |      |       |      |
|     | 5  | 758.225 | 0.154 |      |       |      |
|     | 6  | 772.537 | 0.033 |      |       |      |
| VII | 0  | 639.490 | 0.022 | 0.37 | 15.61 | 0.04 |
|     | 3  | 686.184 | 0.023 |      |       |      |
|     | 4  | 702.226 | 0.014 |      |       |      |
|     | 5  | 718.260 | 0.019 |      |       |      |
|     | 6  | 733.743 | 0.027 |      |       |      |
|     | 7  | 748.411 | 0.026 |      |       |      |
|     | 8  | 764.116 | 0.038 |      |       |      |
|     | **9**  | **780.078** | **0.019** |      |       |      |
|     | 10 | 795.897 | 0.011 |      |       |      |

**Table 3S. Final transit and anomaly phases for each of the largest spots occulted by planet Kepler 30c.**

| Kepler Transit # | $\phi_{anom}$ [deg] | Error | $\phi_{tra}$ [deg] | Error |
|---|---|---|---|---|
| 0 | 15 | 2 | 22 | 10 |
| 2 | 59 | 9 | 34 | 9 |
| 3 | -39 | 12 | -36 | 9 |
| 5 | 48 | 7 | 54 | 3 |
| 10 | 5 | 3 | 1 | 8 |

**Table 4S. Dynamical fit to Transit Times and Durations (Table 1S) and Impact Parameters (Table 1).**

| planet | P (days) | $T_0$ (BJD-2454900) | e cos $\omega$ | e sin $\omega$ | i (deg) | $\Omega$ (deg) | $M_p/M_*$ (x$10^{-6}$) |
|---|---|---|---|---|---|---|---|
| b | **29.33434** | **346.6476** | **0.03616** | **-0.02204** | **90.179** | **0.035** | **34.29** |
| +/- | 0.00815 | 0.0401 | 0.00185 | 0.00638 | 0.167 | 0.167 | 3.03 |
| c | **60.323105** | **357.887042** | **0.00728** | **-0.008332** | **90.3227** | **0.00** | **1935** |
| +/- | 0.000244 | 0.000520 | 0.00133 | 0.000767 | 0.0302 | (def) | 167 |
| d | **143.34394** | **373.53020** | **-0.02060** | **-0.00635** | **89.8406** | **1.319** | **70.09** |
| +/- | 0.00858 | 0.00969 | 0.00510 | 0.00239 | 0.0202 | 0.475 | 5.76 |

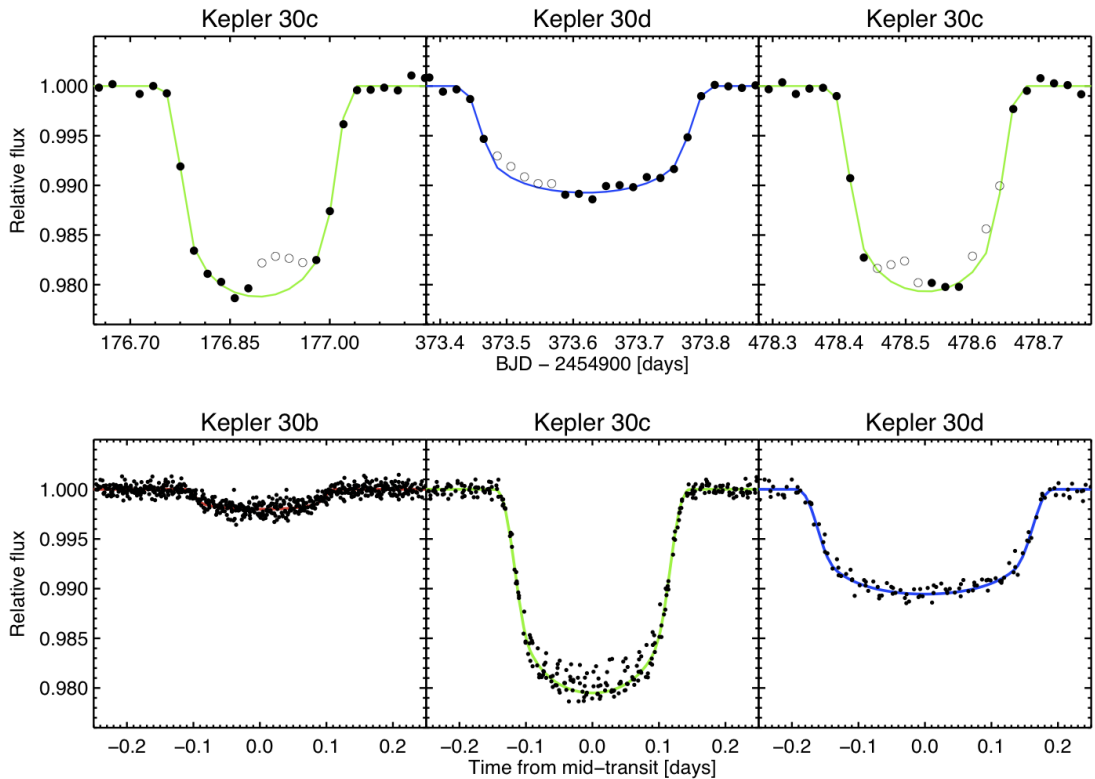

**Figure 1s. Transit curve analysis allowed us to determine the orbital parameters and also the sizes of the planets, properly taking into account the effect spots.**

The upper panel shows three different transits in which spot anomalies are observed. The solid dots represent the observed fluxes used to determine the transit parameters. The open dots represent the observed fluxes affected by spot-crossing events, points that were not used in the transit analysis. The line represents the final transit model that fits through the solid dots.

The lower panel shows the folded light curve for the three planets in which the solid dots represent all observations and the lines represent the final transit model. The effect of the spots seems to be present for the three planets, but it becomes much more evident for Kepler 30c, the largest planet.

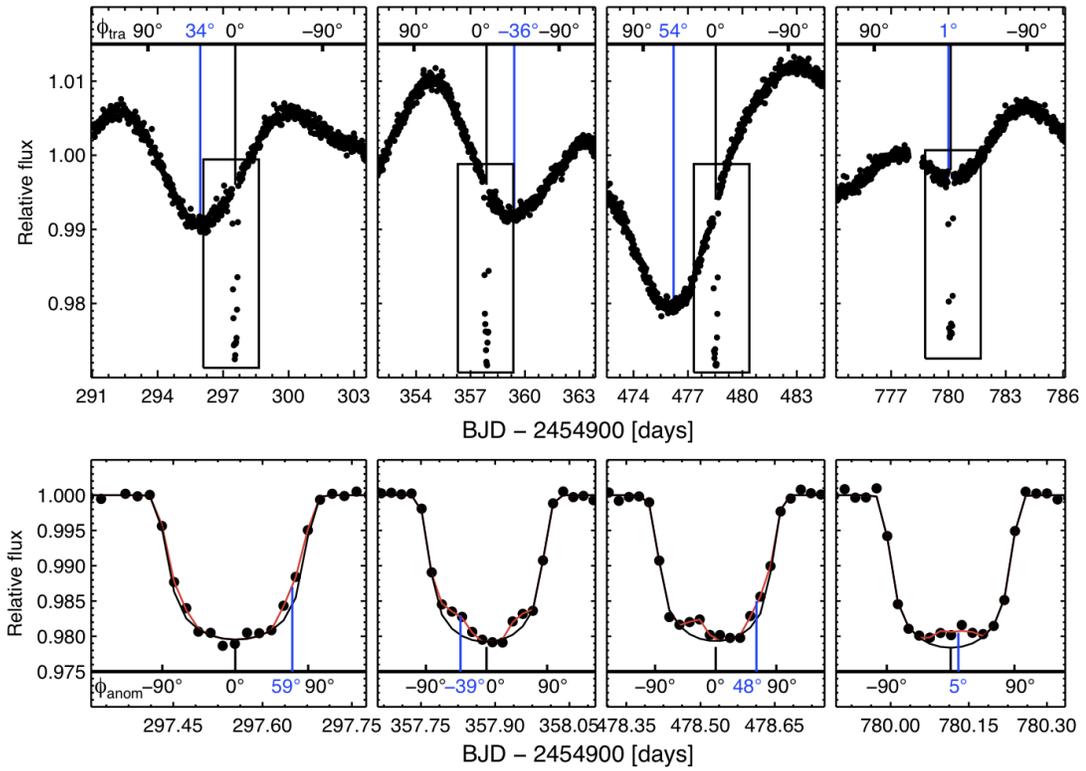

**Figure 2s: Continuation of Figure 1, the transit phases and anomaly phases for the four other spot-crossing events.**

The upper panels are the equivalent of Figure 1a, the lower panels the equivalent of Figure 1b, for all four other spot-crossing events. It is important to note that except for the one on the right side, the other three are based in a model with two spots on the transit chord. In those cases, only one out of the two anomalies happens to be caused by a large enough spot, and that is the one connected with the blue vertical line on the lower panels. See table 2S and 3S for more information.